\documentclass[12pt]{iopart}
\usepackage{graphicx}

\begin{document}

\title[Inflection point as a characteristic of the QCD critical point]{Inflection point as a characteristic of the QCD critical point}

\author{Mingmei Xu and Yuanfang Wu}

\address{Key Laboratory of Quark and Lepton Physics (MOE) and Institute of Particle
Physics, Central China Normal University, Wuhan 430079, China}
\ead{xumm@iopp.ccnu.edu.cn}

\begin{abstract}
The appearance of the inflection point in the equation of state
(EoS) is associated with the second order phase transition. The high
cumulants of conserved quantities near the critical point are
corresponding to the high derivatives of the EoS near the inflection
point. The critical behavior of  high cumulants of conserved charge
near the QCD critical point, in particular, the sign change,  is closely
related to the appearance of inflection point. We show in general
how the times of sign change of high cumulants relate to the order
of derivative. We also demonstrate that the character of inflection
point of EoS is as visible as the sign change of high cumulants in 3
systems, i.e., van der Waals equation of fluid, magnetization of spin
models and the baryon number density of QCD matter. Therefore, we
propose that the EoS, or the mean of baryon number density, should
be measured and studied together with its higher cumulants in
exploring QCD critical point in heavy ion collisions.
\end{abstract}

\maketitle
\section{Introduction}
\def\tc{T_{\rm c}}
\def\d{{\rm d}}

The theory of strong interaction---quantum chromodynamics
(QCD)---has a complicated phase structure. Mapping the QCD phase
diagram in the temperature ($T$) and baryon chemical potential
($\mu_{\rm B}$) plane is currently one of the main goals of high
energy nuclear physics. Lattice QCD calculations indicate that the
chiral and deconfinement phase transitions are a smooth crossover at
zero baryon chemical potential~\cite{crossover}, while several
QCD-based models predict a first order phase transition at high
density~\cite{first-order-1,first-order-2,first-order-3,first-order-4,first-order-5}.
The existence of the QCD critical point (CP), which terminates the
first order phase transition line in the QCD phase diagram, is
expected and being searched for in the ongoing heavy ion
experiments~\cite{review-CP-search}.

Locating the QCD critical point is a challenge for both the
theorists and the experimentalists. Due to that the precise position
of the critical point is not well known from the theoretical side, a
measurement from the experimental side is highly needed. By varying
the center-of-mass energy $\sqrt{s_{\rm NN}}$ of the nucleus-nucleus
collisions, we can scan large regions of the phase diagram. The
chemical freeze-out at different $\sqrt{s_{\rm NN}}$ happens at
different positions along the freeze-out
curve~\cite{chemical-freezeout} and therefore the trajectories of
the reaction systems in the $T$-$\mu_{\rm B}$ plane have to cross
different areas of the phase diagram and might even hit the critical
area. When the trajectories of the reaction systems in the
$T$-$\mu_{\rm B}$ plane get close to the critical point, large
fluctuations appear. Assuming freeze-out happens near the CP, the
large fluctuations can survive and should be observed. A
non-monotonic behavior of the fluctuations from low energy to high
energy is expected~\cite{propose-non-monotonic}. Several
event-by-event fluctuation observables, e.g. mean transverse
momentum fluctuations and the multiplicity fluctuations (here the
fluctuation measures relate to the variances of the event-variable
distributions, or the second order cumulants), have been analyzed at
the SPS energies with $5<\sqrt{s_{\rm
NN}}<17$~GeV~\cite{variance-NA49} and at the RHIC energies with
$7<\sqrt{s_{\rm NN}}<200$~GeV~\cite{variance-STAR}. NA49 at SPS finds
peaks in the system size dependence of the mean transverse momentum
fluctuation and the multiplicity fluctuation, which agrees with
predictions for the CP. However, the energy dependences of the two
measures do not show any signatures for CP. The evidences for a
critical point are inconclusive so far~\cite{review-variance}.

The critical point means the divergent correlation length in the
thermodynamic limit. While in a realistic heavy ion collision the
correlation length $\xi$ gets at most $2$-$3$~fm because of the
finite system size and the limited evolution
time~\cite{limited-correlation-length}. Since higher cumulants are
proportional to a higher power of $\xi$, higher cumulants of
conserved charges are more sensitive to the critical point and were
proposed as a promising observable for the search of the QCD
critical
point~\cite{propose-high-moment-1,propose-high-moment-2,propose-high-moment-3}.
The cumulants are calculated from corresponding probability
distributions in experiments. Theoretically, according to the grand
canonical formulation of thermodynamics, the cumulants for charge X
are proportional to generalized susceptibilities, which are
derivatives of the pressure with respect to the corresponding
Lagrange multiplier $\mu_{\rm X}$, i.e.
\begin{equation}
\chi^{\rm
X}_{n}=\frac{\partial^{n}\widehat{p}}{\partial\widehat{\mu}^{n}_{\rm
X}},\label{derivative1}
\end{equation}
where $\widehat{\mu}_{\rm X}=\mu_{\rm X}/T$ and $\widehat{p}=P/T^4$.
Since the baryon number density is $\langle n_{\rm
B}\rangle=\frac{\partial P}{\partial\mu_{\rm B}}$, the $n$th order
cumulant is related to the $(n-1)$th order derivative of $\langle
n_{\rm B}\rangle$ with respect to $\mu_{\rm B}$, i.e.
\begin{equation} \chi^{\rm B}_{n}\sim \frac{\partial^{n-1}\langle n_{\rm B}\rangle}{\partial\mu^{n-1}_{\rm
B}}. \label{chi-nB}
\end{equation} Besides the cumulants of charges, the cumulants of the energy
are also proposed as probes of the QCD phase
transition~\cite{Asakawa,Koch-talk}. At $\mu=0$, the cumulants of
the energy measure the derivatives of the EoS~\cite{Koch-talk}, i.e.
\begin{equation}
\langle(\delta E)^n\rangle=(-\frac{\partial}{\partial
1/T})^{n-1}\langle E\rangle.\label{derivative2}
\end{equation}

Mostly the signals are characterized by the sign or the change of
the sign of various cumulants. For example, based on a general
analysis and the NJL model calculations, the sign changes of the
third order cumulants are proposed as signals of passing across the
phase boundary~\cite{Asakawa}. Based on an universal analysis and
the non-linear sigma model calculations, the fourth order cumulants
(kurtosis) should be negative when the critical point is approached
on the crossover side~\cite{Stephanov} and will change signs in the
critical region. The possible change in the sign of the kurtosis in
the critical region is also pointed out basing on a lattice
calculation~\cite{Gupta}. The oscillating behavior of the sixth
order cumulants $\chi^{B}_{6}$ (or correspondingly the ratio of
cumulants $R^{\rm B}_{6,2}=\chi^{\rm B}_{6}/\chi^{\rm B}_{2}$) and
the eighth order cumulants $\chi^{B}_{8}$ (or $R^{\rm
B}_{8,2}=\chi^{\rm B}_{8}/\chi^{\rm B}_{2}$) show up in the
crossover region, according to the analysis of the universal scaling
functions, the PQM models~\cite{Friman,Skokov} and the lattice
calculations~\cite{propose-high-moment-3,lattice-chi6}. However, due
to the difficulties of the lattice calculations and the model
estimations, the high cumulants of the net baryon number are still
not final. The experimental measurements of high cumulants have
various difficulties either, e.g. the centrality bin width
effect~\cite{CBWC}, the lack of statistics especially at low
energies, etc. Further analysis methods are being developed.

Since the cumulants of conserved charges are related to the
derivatives of the equation of state, as shown in equations
(\ref{chi-nB}) and (\ref{derivative2}), oscillating high moments
mean oscillating high derivatives. As the oscillation behavior or
the sign change behavior is referred to in many papers, we have to
ask: which feature of the primary function (EoS) does the
oscillation of high derivatives reflect? How the critical point
manifests itself in the equation of state?

In this respect, we notice that all of the observables mentioned
above are related to the divergent characteristic quantities in a
second order phase transition. For example, the cumulants of the
energy are related to the heat capacity at fixed volume as
\begin{equation}
C_{V}=(\frac{\partial U}{\partial T})_V, \end{equation}
 with the
internal energy $U=\langle E\rangle$. If we call $\langle
E\rangle(T)$ primary function (it can be regarded as the EoS), the
second order cumulant in (\ref{derivative2}) is proportional to
$C_{V}$ and higher cumulants are related to the derivatives of
$C_{V}$. The divergent peak of $C_{V}$ will result in the sign
change of the next order derivative. Since $C_{V}=(\frac{\partial
U}{\partial T})_V=T(\frac{\partial S}{\partial T})_V$, a pair of
conjugate variables, $S$ and $T$, are involved. Besides, the
divergent compression coefficient
\begin{equation}\kappa_T\equiv-\frac{1}{V}(\frac{\partial V}{\partial
p})_T,\label{kappa}\end{equation}
 involves another pair of conjugate variables, $V$
and $p$. The divergent susceptibility
\begin{equation}\chi=(\frac{\partial M}{\partial H})_T
\end{equation} in the ferromagnetic transition involves conjugate variables $M$
and $H$. The divergent generalized susceptibility in equation
(\ref{chi-nB}) involves conjugate variables $N_{\rm B}$ and
$\mu_{\rm B}$. All of these divergent quantities ($C_{V}$,
$\kappa_T$, $\chi$ and $\chi_2^{\rm B}$) are the derivatives of an
extensive variable ($S$, $V$, $M$ and $N_{\rm B}$) with respect to
the conjugate intensive variable ($T$, $p$, $H$ and $\mu_{\rm B}$).
If we call $S(T)$(or $U(T)$), $V(p)$, $M(H)$ and $N_{\rm B}(\mu_{\rm
B})$ as the equations of state, which feature will result in the
sign change or oscillation in their high derivatives?

In this paper, we use a simple
curve with an inflection point as an example and obtain oscillating high order derivatives,
which is shown in Section~\ref{func-osci}. The law about how the number
of times of sign change in the neighborhood of the inflection point relates to the order of derivatives
is also summarized. In Section~\ref{three-sys} by analyzing the
inflection point of van der Waals equation of fluid, magnetization of spin models and the baryon number density of QCD matter, we demonstrate that the character of inflection point of EoS is as visible as the sign change of high cumulants.  In
Section~\ref{conclusion} we propose to measure the EoS, or the mean of the baryon number density together with its higher cumulants in exploring QCD critical point in heavy ion collisions.

\section{Inflection point and the oscillations of the higher derivatives}\label{func-osci} An inflection point is defined as a
point on a curve at which the sign of the concavity changes. Since
the concavity can be reflected by the sign of the second derivative,
an inflection point also means a point on a curve at which the
second derivative changes sign. So the second derivative is helpful
to judge an inflection point. A necessary condition for $x$ to be an
inflection point is the second derivative, $f''(x)$, is equal to
zero if it exists. A sufficient condition requires $f''(x+\epsilon)$
and $f''(x-\epsilon)$ to have opposite signs in the neighborhood of
$x$.

\begin{figure}[!htp]
\centering
\begin{minipage}[c]{0.3\textwidth}
\centering
\includegraphics[width=\textwidth]{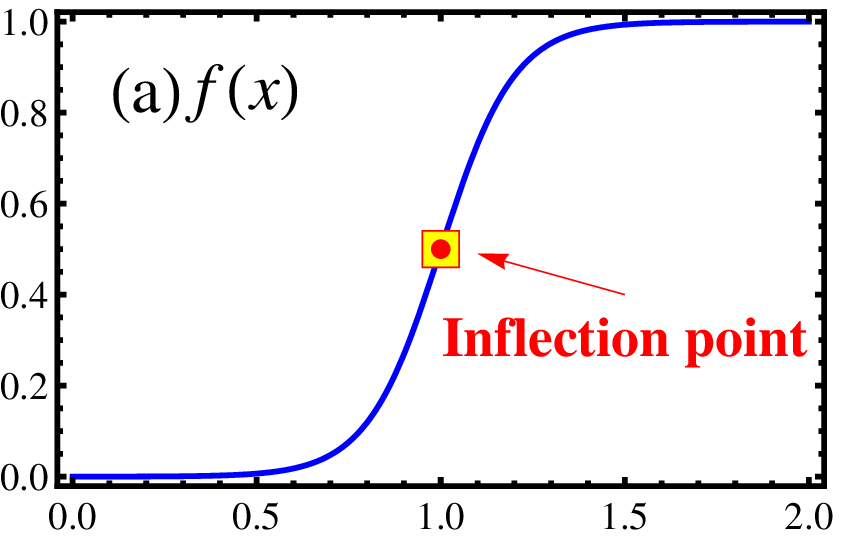}
\end{minipage}%
\hskip 0.2cm
\begin{minipage}[c]{0.3\textwidth}
\centering
\includegraphics[width=\textwidth]{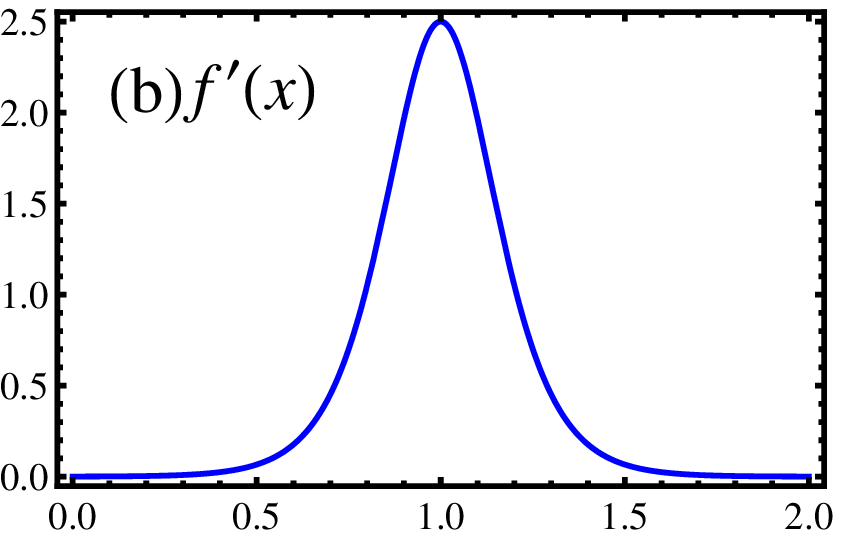}
\end{minipage}
\hskip 0.2cm
\begin{minipage}[c]{0.3\textwidth}
\centering
\includegraphics[width=\textwidth]{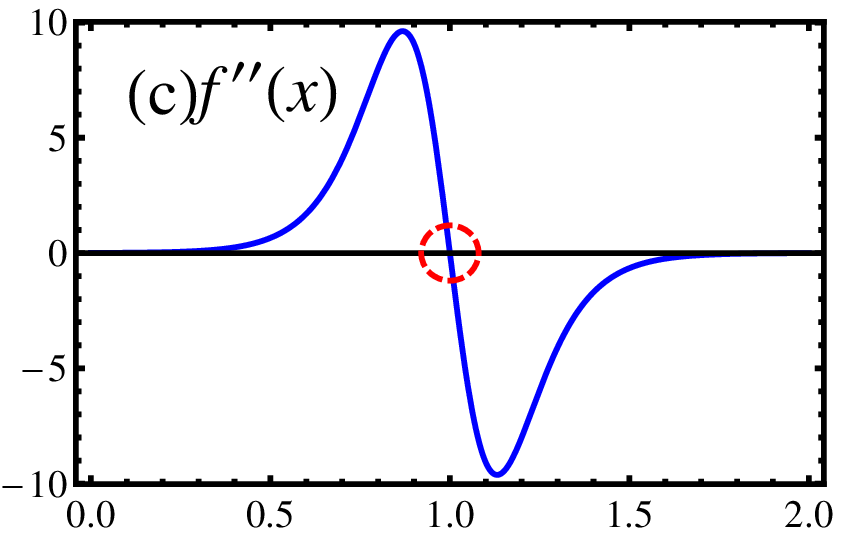}
\end{minipage}\\
\vskip 0.2cm
\begin{minipage}[c]{0.3\textwidth}
\centering
\includegraphics[width=\textwidth]{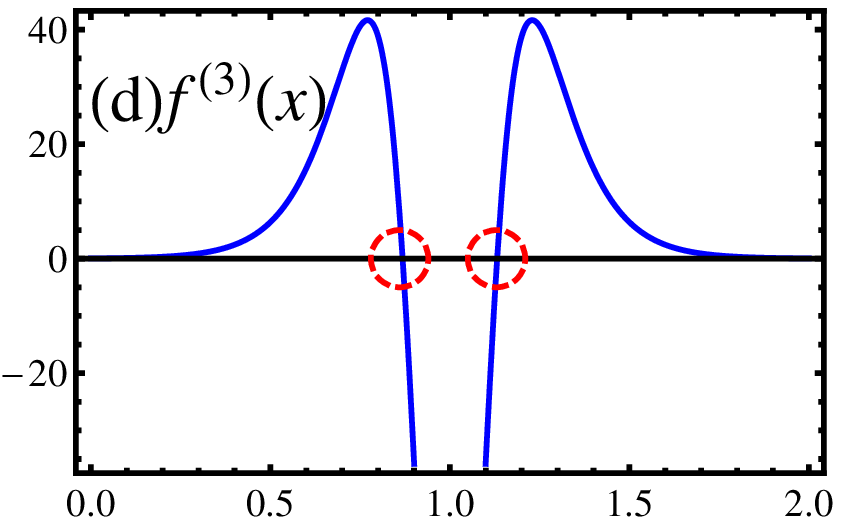}
\end{minipage}%
\hskip 0.2cm
\begin{minipage}[c]{0.3\textwidth}
\centering
\includegraphics[width=\textwidth]{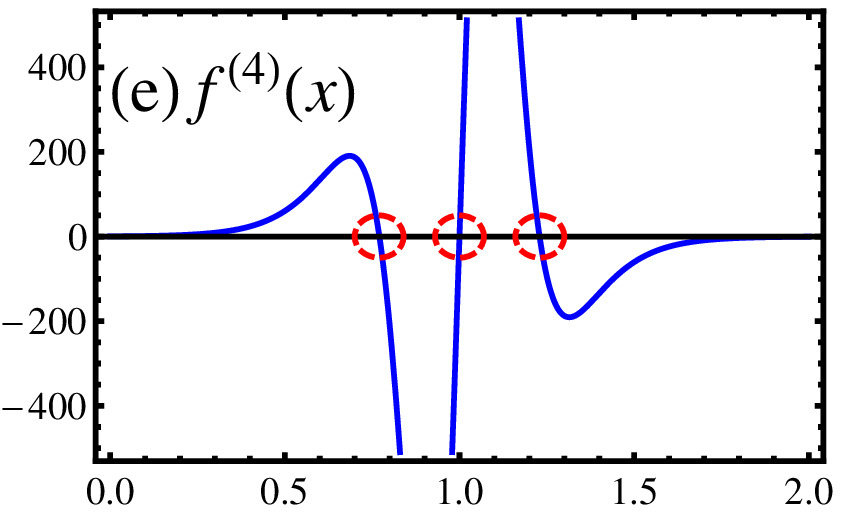}
\end{minipage}
\hskip 0.2cm
\begin{minipage}[c]{0.3\textwidth}
\centering
\includegraphics[width=\textwidth]{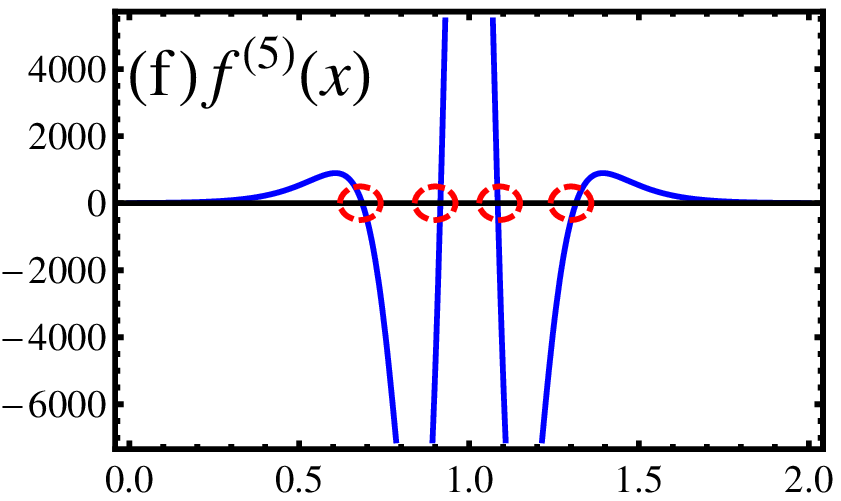}
\end{minipage}
\caption{(a) The curve of the function $f(x)$ in equation
(\ref{ex-func}) which has an inflection point at $x=1$, marked as
red point in the subfigure. (b)-(f) Its $n$th order derivatives from
$n=1$ to $n=5$, with the red open circles denoting the positions of
sign change. \label{figure1}}
\end{figure}

 Let us demonstrate using a function with an inflection point, e.g.
\begin{equation}
f(x)=\frac{1+\tanh[5(x-1)]}{2}.\label{ex-func}
\end{equation} Through requiring
$f''(x)=0$ and observing the signs of $f''(x-\epsilon)$ and
$f''(x+\epsilon)$, we find that point $(1,0.5)$ is an inflection
point, as shown in figure~\ref{figure1}(a). As $x$ increases from 0
to 2, the curve changes from concave to convex. The inflection point
(marked as red point in figure~\ref{figure1}(a)) is the point where
the concavity changes. Then we plot the derivatives from the first
order to the fifth order and show them in
figure~\ref{figure1}(b)-(f). The first derivative (the slope of the
curve) shows a peak on the interval and is at an extremum at the
inflection point, i.e. a (local) minimum or maximum. At the
inflection point, the second derivative changes the sign. The
positions of the sign change are denoted by red open circles in
figure~\ref{figure1}. We see oscillating behavior of the high
derivatives. In the neighborhood of the inflection point, the third
order derivative changes the sign two times, the fourth order
derivative changes the sign three times, and the fifth order
derivative changes the sign four times. From this example function
(\ref{ex-func}), we summarize to obtain the law about the number of
times of sign change (the counting law): the $n$th order derivative changes the sign
$(n-1)$ times in the neighborhood of the inflection point.

\section{Inflection point of EoS and the critical
point}\label{three-sys}

How the inflection point relates to the critical point are interesting in general and in particular, in QCD system. So we choose three systems in the section to see their relations.

\subsection{Inflection point of the van der Waals equation of fluid\label{IP-liquid}}
The van der Waals equation is an equation of state for a fluid which
qualitatively describes the properties of the liquid phase and the
gas phase as well as the phase transition between them. Although the
van der Waals equation can not describe the coexistence state at the
first order phase transition, it is still a good tool to demonstrate
a critical point. The equation reads
\begin{equation}
(p+a/v^2)(v-b)=RT,
\end{equation}
with $p, v, T$ representing the pressure of the fluid, the molar
volume and the temperature respectively. $a$ and $b$ are parameters
which depends on the type of the molecule. $R$ is the universal gas
constant $8.2057\times10^{-2}$
atm$\cdot$l$\cdot$mol$^{-1}\cdot$K$^{-1}$.

\begin{figure}[!htp]
\centering
\begin{minipage}[c]{0.33\textwidth}
\centering
\includegraphics[width=\textwidth]{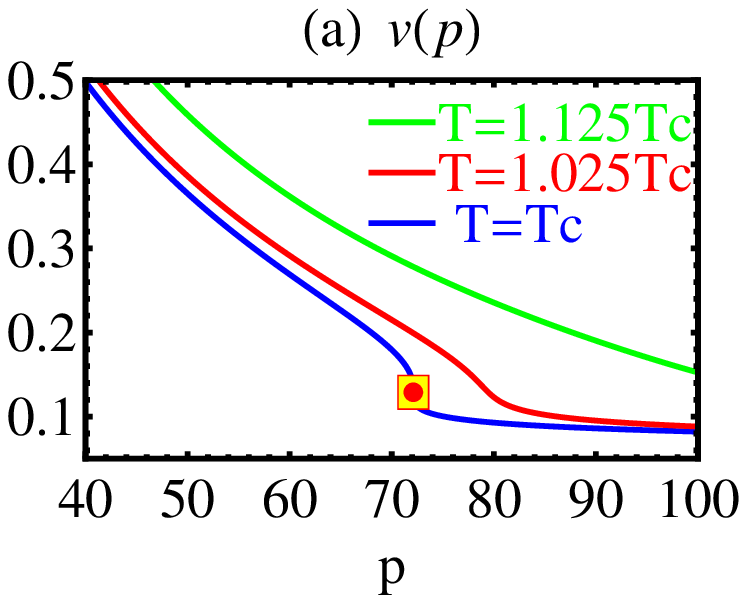}
\end{minipage}%
\hskip -0.5cm
\begin{minipage}[c]{0.35\textwidth}
\centering
\includegraphics[width=\textwidth]{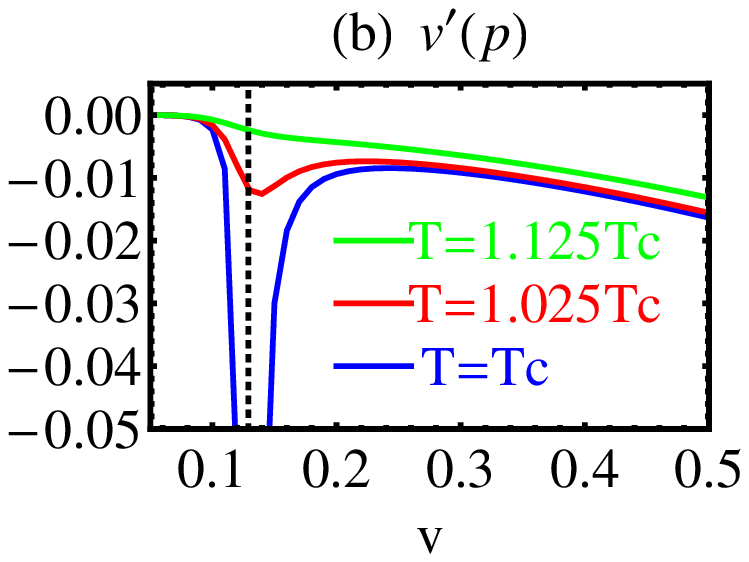}
\end{minipage}
\hskip -0.5cm
\begin{minipage}[c]{0.35\textwidth}
\centering
\includegraphics[width=\textwidth]{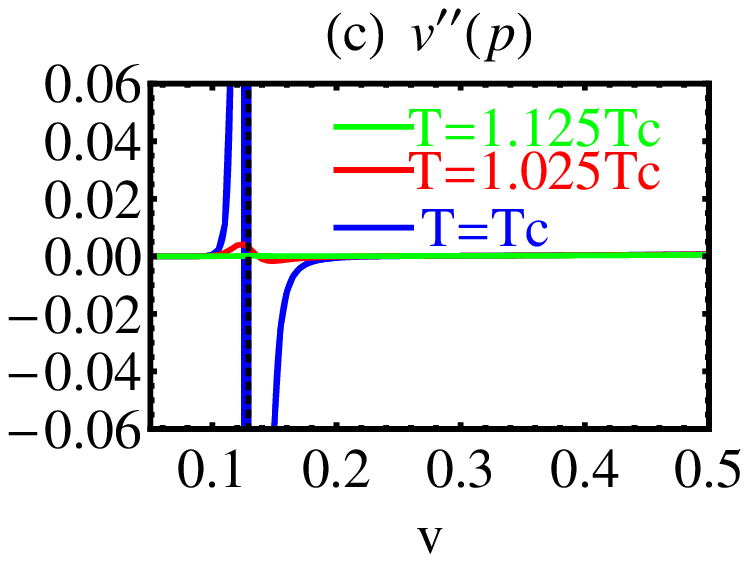}
\end{minipage}
\caption{(a) The van der Waals isotherms ($v$-$p$ plot) for $T_c$ in
blue, $1.025T_c$(=310 K) in red and $1.125T_c$(=340 K) in green. An
inflection point is marked as red point in the critical isotherms.
(b) The first derivative $v'(p)$ at fixed temperature. The black
dashed line is to guide us the position of the critical point. (c)
The second derivative $v''(p)$ at fixed temperature.
\label{figure2}}
\end{figure}

For demenstrating, here we use the parameters of CO$_2$, i.e.
$a=3.6$ atm$\cdot$l$^2\cdot$mol$^{-2}$ and $b=0.043$
l$\cdot$mol$^{-1}$. Its critical temperature, critical volume and
critical pressure would be $T_c=302.305$ K, $v_c=0.129$ l,
$p_c=72.111$ atm. The van der Waals isotherms ($v$-$p$ plot) for
$T_c$, $1.025T_c$(=310 K) and $1.125T_c$(=340 K), are shown in
figure~\ref{figure2}(a). We do not show the isotherms for $T<T_c$
since experimental isotherms do not follow the van der Waals curves
at that case, as we mentioned above. When $T=T_c$, the blue curve
shows an inflection point, as marked in red in
figure~\ref{figure2}(a). As the temperature increases, the
inflection point disappears gradually and the curve turns to concave
in the whole interval.

The first derivative of $v(p)$ at fixed temperatures is shown in
figure~\ref{figure2}(b). When $T=T_c$, the first derivative
$(\frac{\partial v}{\partial p})_T$ shows a cuspal peak at the
critical position. As the temperature increases to $1.025T_c$, the
peak shrinks. At $1.125T_c$, the peak nearly disappears. The first
derivative of $v(p)$ at fixed temperatures is proportional to the
isothermal compression coefficient (see equation (\ref{kappa})). The
discontinuity or divergence of the compression coefficient is one of
the features of the second order phase transition passing through
the critical point. The divergence of $\kappa_T$ is equivalent to
the divergence of the first derivative $(\frac{\partial v}{\partial
p})_T$. Here we show that it is the inflection point in
figure~\ref{figure2}(a) that leads to the divergence of the
compression coefficient.

Figure~\ref{figure2}(c) shows the second derivative of $v(p)$ at
fixed temperatures. At $T=T_c$, prominent sign change occurs at the
critical position, which exactly is the inflection point in the
critical isotherms in figure~\ref{figure2}(a). An inflection point
certainly means the sign change of the second derivative, as
defined. In paper~\cite{Asakawa}, the sign change is proposed as the
signal of the critical point. We note that, the intrinsic
nature of the sign change is the existence of an inflection point in
the equation of state. Essentially it is the inflection point that
leads to the sign change of high order derivatives. In a broad
sense, inflection point of EoS is a characteristic of the critical
point.

It is worth noting that the derivatives of $v$ with respect to $p$
show signals of the CP, while the converse of this derivatives, i.e.
the derivatives of $p$ with respect to $v$, do not show any signals.
This means that we may not observe the critical behavior by improper
observables. The selected observables greatly affect whether we
observe the CP. According to an analysis in the introduction, we
regard the derivatives of the extensive variable with respect to its
conjugate intensive variable as good observables.

\subsection{Inflection point of the magnetization in the spin models}

\begin{figure}[!htp]
\centering
\begin{minipage}[c]{0.33\textwidth}
\centering
\includegraphics[width=\textwidth]{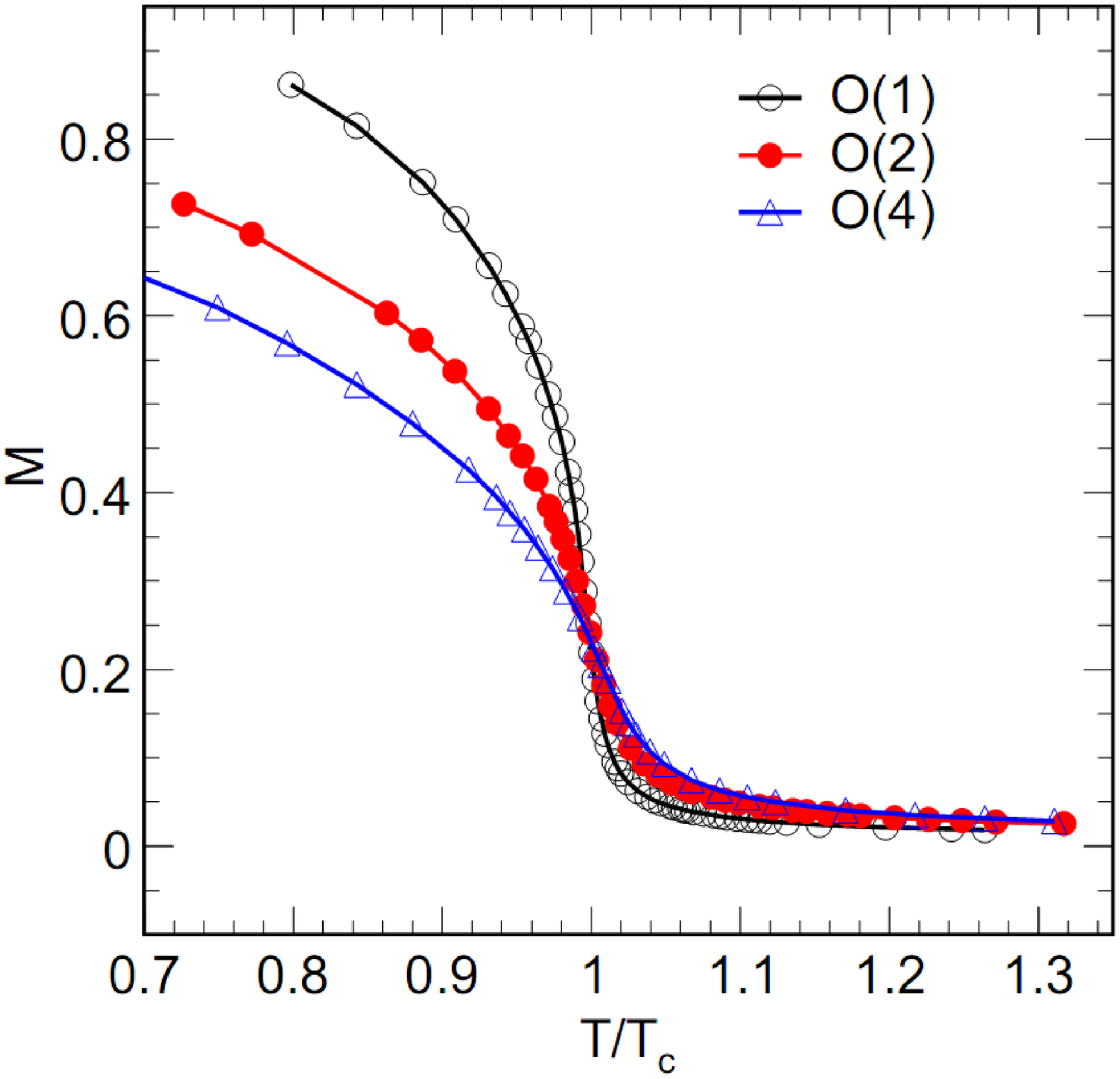}
\put(-30,70){\small (a)}
\end{minipage}%
\begin{minipage}[c]{0.33\textwidth}
\centering
\includegraphics[width=\textwidth]{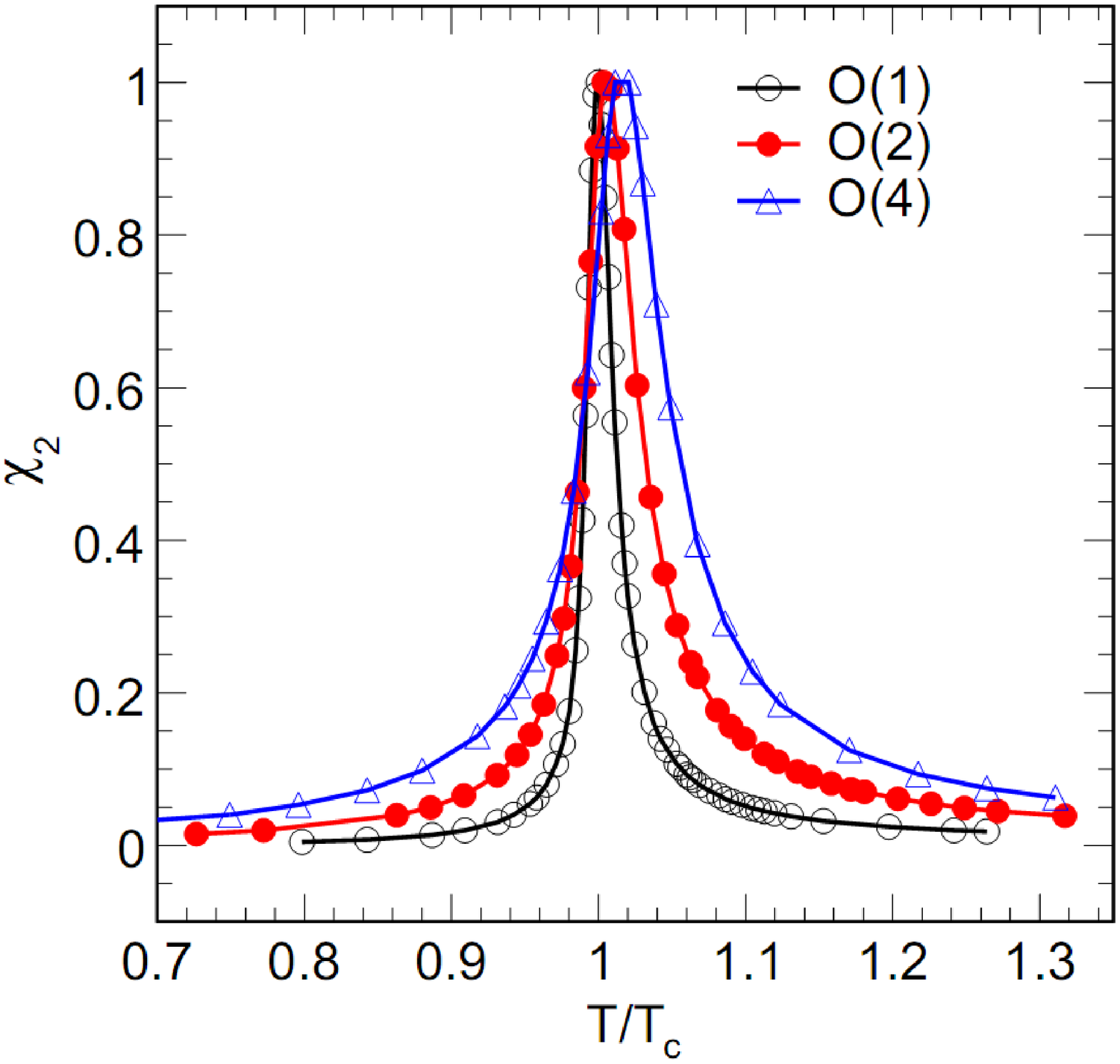}
\put(-30,70){\small (b)}
\end{minipage}
\begin{minipage}[c]{0.33\textwidth}
\centering
\includegraphics[width=\textwidth]{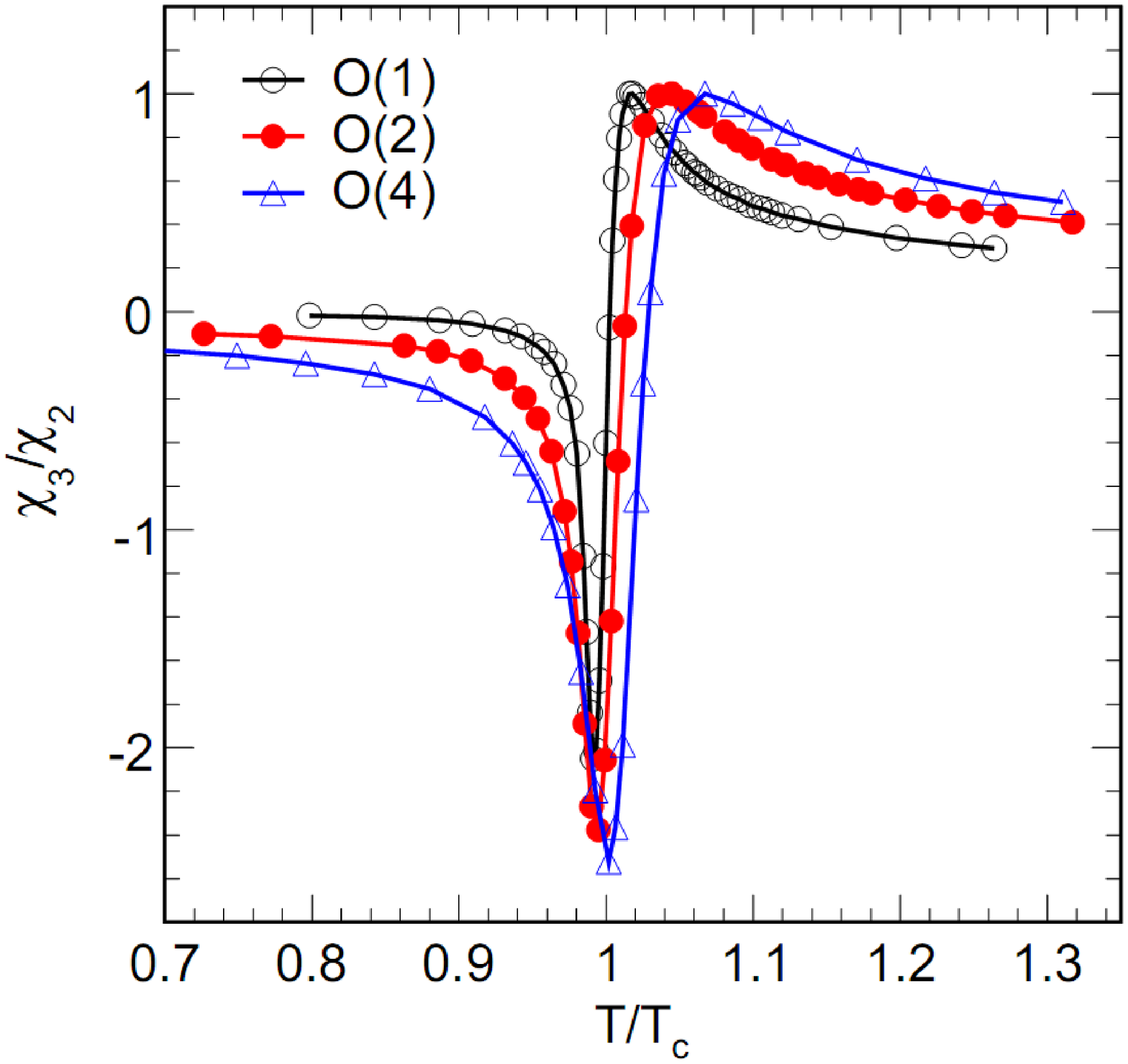}
\put(-30,70){\small (c)}
\end{minipage}
\caption{(a) The first order cumulant $\chi_1$ defined in equation
(\ref{chi-spin}), i.e. the magnetization $M$, (b) the quadratic
cumulant $\chi_2$ defined in equation (\ref{chi-spin}), i.e. the
first derivative of $M$, (c) the cubic cumulant$\chi_3$, i.e. the
second derivative of $M$, for Ising (O(1)) (black points), O(2) (red
points) and O(4) (blue points) models. An inflection point is
obviously seen in the subfigure (a). The figures are from
reference~\cite{PanX}. \label{figure3}}
\end{figure}

In the 3-dimensional O(1)(Ising), O(2) and O(4) spin models without
external magnetic field, which are good examples to show the
critical behaviors, it is found that all higher cumulant ratios
change dramatically the sign near the critical
temperature~\cite{PanX}. They define
\begin{equation}
\chi_n=-(\frac{\partial^n f}{\partial H^n})_T,\label{chi-spin}
\end{equation}
where $f$ denotes the free energy density and $H$ denotes the
magnitude of the external magnetic field. The first derivative in
equation (\ref{chi-spin}), $\chi_1=-(\frac{\partial f}{\partial
H})_T$, is the magnetization $M$, i.e. the order parameter in the
spin models, which is shown in figure~\ref{figure3}(a). The function
$M(H,T)$ is usually regarded as the equation of state.
(\ref{chi-spin}) is equivalent to
\begin{equation} \chi_n=(\frac{\partial^{n-1} M}{\partial
H^{n-1}})_T.\label{chi-M}
\end{equation} An inflection point is obviously seen in figure~\ref{figure3}(a). According to
equation (\ref{chi-M}), $\chi_2=(\frac{\partial M}{\partial H})_T$
is the usual susceptibility, which shows a finite peak in the
vicinity of critical temperature in the Monte Carlo simulation, see
figure~\ref{figure3}(b). The magnetization process without external
magnetic field, as simulated in~\cite{PanX}, is a second order phase
transition. The susceptibility should be divergent in the
thermodynamic limit and be a finite peak in a finite system, as
expected. The second derivative of EoS $(\frac{\partial^2
M}{\partial H^2})_T$, i.e. $\chi_3$ plotted in
figure~\ref{figure3}(c), shows sign change. The sign change in the
second derivative means there is an inflection point in the EoS.
They also report oscillating behavior in $\chi_4$ and $\chi_6$
in~\cite{PanX} which are not shown here. Essentially, it is the
inflection point of EoS that results in an oscillating behavior of
the high cumulants.

We have noticed that the inflection point for a finite system is not
exactly the critical point, i.e., where the sign of the second
derivative changes is above the critical temperature $\tc$, as
figure~\ref{figure3}(c) shows. For a finite system, inflection point
is just an approximate value of the critical point. As the system
size increases, the inflection point will approach the critical
point, which was previously discussed in a percolation
model~\cite{kehw-percolation}.

\subsection{Inflection point of the baryon number density of QCD matter\label{IP-crossover}}

Though the transition from hadron phase to quark gluon plasma phase
is a crossover type at vanishing baryon chemical
potential~\cite{crossover}, the critical behavior can also be
observed during a
crossver~\cite{Friman,Redlich-CPOD2011,Skokov-CPOD2011} due to the
remnants of O(4) criticality on the crossover line. The high
cumulants $\chi_4^{\rm B}/\chi_2^{\rm B}$, $\chi_6^{\rm
B}/\chi_2^{\rm B}$ and $\chi_8^{\rm B}/\chi_2^{\rm B}$ from an
universality analysis and the PQM model calculations, indeed show
oscillating behavior close to the pseudo-critical temperature (the
temperature of crossover). This oscillating behavior of high
cumulants of the baryon number should be related to the inflection
point of the baryon number density, as equation (\ref{chi-nB})
suggests. Inflection points indeed exist in the curves of the quark
number density $n_{\rm q}(\mu_{\rm q})$ for $T=\tc$ and $T>\tc$, as
shown in an earlier paper~\cite{Schaefer-PRD2007}. The quark number
density is just the baryon number density except a factor of $1/3$.

To summarize this section, all the existing second order phase
transitions show the characteristic of an inflection point. An
inflection point of EoS is a characteristic of the critical point
and is associated with the oscillating behaviors of higher
cumulants. Physically, the second order phase transition passing
through the CP is always accompanied by the numerical change of the
order parameter. Assuming that the temperature is a controlling
parameter. The order parameter, denoted by $O$, gets a particular
value (e.g. $O_{a}$) at the low temperature limit in one phase and
another value (e.g. $O_{b}$) at the high temperature limit in
another phase. During a second order phase transition, the value of
the order parameter is transformed continuously. When the
temperature increases from a low value, the order parameter will
deviate from $O_{a}$ and the curve of $O(T)$ at this stage would be
concave. When the temperature is high enough, the order parameter
will turn to be convex to approach its upper limit $O_{b}$. The
point where the concavity changes is an inflection point. Therefore,
a second order phase transition certainly means an inflection point
on the equation of state. In this sense, inflection point of EoS is
a characteristic of the critical point. This statement is general
and independent on any models.

In section~\ref{IP-crossover}, an inflection point also appears in
the EoS of crossover. In fact, the inflection points of some
variables have been used to determine the transition temperature of
crossover at $\mu_{\rm B}=0$ in lattice calculations
~\cite{lattice-Tpc-01,lattice-Tpc-02,lattice-Tpc-03}. The fact that
inflection point appears in the EoS of crossover does not contradict
the statement in the previous paragraph, however. As we showed in
section~\ref{IP-liquid}, when the crossover happens near the CP,
e.g. $T=1.025\tc$, an inflection point survives, the first
derivative shows a peak and the second derivative changes the sign.
As the temperature increases, the peak in the first derivative
shrinks and the sign change in the second derivative is not obvious.
As the temperature increases further, the peak and sign change
behavior disappear at last. The inflection point in the EoS of
crossover may be due to the close location with the CP. This trend
is common for different systems. For example, in the
three-dimensional three-state Potts model in an external
field~\cite{Karsch-Potss}, the peak heights of the susceptibilities
decrease as the external field $h$ increases away from the critical
point to crossover region (see figure 1(a) in~\cite{Karsch-Potss}).
Similarly, all the baryon number susceptibilities (or the quark
number susceptibilities) calculated in NJL
model~\cite{Asakawa,chi-NJL}, another kind of chiral
model~\cite{Hatta} and the lattice
QCD~\cite{chiq-lattice-01,chiq-lattice-02,chiq-lattice-03} show the
same trend, too. Since the peak heights of the susceptibilities
($\chi_2$) reflect the slope at the inflection point in the EoS, a
finite peak in a crossover corresponds to a finite value of the
slope at the inflection point in the EoS, while a divergent peak at
the CP corresponds to a infinite value of the slope at the
inflection point in the EoS. As the critical point is approached
from the crossover side, the susceptibility gradually develops to be
divergent and the slope at the inflection point in the EoS gradually
increases to infinity. The fact that the peak structure of the so
called susceptibility survives along the crossover line induces the
phrase ``critical region", as mentioned
in~\cite{propose-high-moment-1,Schaefer-PRD2007}.

The analysis about the inflection point and oscillating high
cumulants is resonable and can be applied. Let us use this analysis
to explain the existing cumulants. It was suggested that the
divergent behavior of $n$th order susceptibility for $\mu_{q}/T>0$
is similar to that of the $2n$th order susceptibility for
$\mu_{q}/T=0$ according to the O(4) universal scaling
analysis~\cite{Friman}. In that universal analysis, for $\mu_{q}/T=0$, $\chi_4^{\rm B}$
shows a peak, $\chi_6^{\rm B}$ changes the sign one time and
$\chi_8^{\rm B}$ changes the sign two times (see figure 1 and 2 in
~\cite{Friman}). The PQM model at $\mu_{q}/T=0$~\cite{Friman}
reproduces the trend very well. In lattice QCD calculations at
$\mu_{q}/T=0$, $\chi_2^{\rm q}$ has an inflection
point~\cite{chiq-lattice-01,chiq-lattice-02,chiq-lattice-03},
$\chi_4^{\rm B}/\chi_2^{\rm B}$ shows a moderate peak and
$\chi_6^{\rm B}/\chi_2^{\rm B}$ changes the sign one
time~\cite{lattice-chi6}. The law about the number of times of sign
change works well. 

\section{Conclusions}\label{conclusion}

In the paper, we firstly show how the critical behavior of high cumulant of susceptibility, in particular,
the sign change, relates to the appearance of inflection point of EoS. The number of times of sign change
of high cumulants near the critical temperature is determined by the order of cumulants, i.e.,
the order of derivative of EoS at the inflection point. The second, forth, sixth
and eighth cumulants of conserved charges at zero chemical potential from current lattice QCD, an O(4) universal scaling analysis as well as PQM model calculation,
show consistent sign changes as we predicted from the order of the cumulants.
Secondly,  we demonstrated that the characters of inflection point of EoS in three systems, i.e.,
Van der Waals fluid, magnetization of spin models and QCD matter,  are as visible as the sign
change of high cumulants of susceptibilities. Therefore, we propose that the EoS, or the mean of
baryon number density, should be measured and studied together with its high cumulants in
exploring QCD critical point in heavy ion collisions.

\ack

This work is supported by the NSFC of china with project nos
11005045, 11221504 and by CCNU-QLPL Innovation Fund (QLPL2011P01).


\def\ea{{\it et al.}}

\end{document}